\documentclass[preprint,review,12pt]{elsarticle}
\usepackage[colorlinks,linkcolor=red,anchorcolor=blue,citecolor=blue]{hyperref}

\usepackage{hyperref}
\usepackage{amssymb}
\usepackage{url}
\newcounter{bla}

\journal{Computer Physics Communications}

\begin{document}

\begin{frontmatter}

%% Title, authors and addresses
\title{{\sc Phasego}: A toolkit for automatic calculation and plot of phase diagram}

\author[a]{Zhong-Li Liu\corref{author}}

\cortext[author] {Corresponding author.\\\textit{E-mail address:} zl.liu@163.com}
\address[a]{College of Physics and Electric Information, Luoyang Normal University, Luoyang 471022, China}

\begin{abstract}
The {\sc Phasego} package extracts the Helmholtz free energy from the phonon density of states obtained by the first-principles calculations. With the help of equation of states fitting, it reduces the Gibbs free energy as a function of pressure/temperature at fixed temperature/pressure. Based on the quasi-harmonic approximation (QHA), it calculates the possible phase boundaries among all the structures of interest and finally plots the phase diagram automatically. For the single phase analysis, {\sc Phasego} can numerically derive many properties, such as the thermal expansion coefficients, the bulk moduli, the heat capacities, the thermal pressures, the Hugoniot pressure-volume-temperature relations, the Gr{\"u}neisen parameters, and the Debye temperatures. In order to check its ability of phase transition analysis, I present here two examples: semiconductor GaN and metallic Fe. In the case of GaN, {\sc Phasego} automatically determined and plotted the phase boundaries among the provided zinc blende (ZB), wurtzite (WZ) and rocksalt (RS) structures. In the case of Fe, the results indicate that at high temperature the electronic thermal excitation free energy corrections considerably alter the phase boundaries among the body-centered cubic (bcc), face-centered cubic (fcc) and hexagonal close-packed (hcp) structures.

\end{abstract}


\end{frontmatter}

%%
%% Start line numbering here if you want
%%
% \linenumbers

% Computer program descriptions should contain the following
% PROGRAM SUMMARY.
\begin{flushleft}
{\bf Program summary}
\end{flushleft}

\begin{small}
\noindent
{\em Manuscript Title:}  {\sc Phasego}: A toolkit for automatic calculation and plot of phase diagram                                    \\
{\em Authors:} Zhong-Li Liu\\
{\em Program Title:} {\sc Phasego}\\
{\em Journal Reference:}                                      \\
  %Leave blank, supplied by Elsevier.
{\em Catalogue identifier:}                                   \\
  %Leave blank, supplied by Elsevier.
{\em Licensing provisions:} GNU GPL version 3\\
  %enter "none" if CPC non-profit use license is sufficient.
{\em Programming language:} Python (versions 2.4 and later)\\
{\em Computer:} Any computer that can run Python (versions 2.4 and later)\\
  %Computer(s) for which program has been designed.
{\em Operating system:} Any operating system that can run Python\\
  %Operating system(s) for which program has been designed.
{\em RAM:} 10 M bytes\\
  %RAM in bytes required to execute program with typical data.
{\em Number of processors used:}                              \\
  %If more than one processor.
{\em Supplementary material:}                                 \\
  % Fill in if necessary, otherwise leave out.
  % Please give some freely chosen keywords that we can use in a
  % cumulative keyword index.
{\em Classification:}  7.8 Structure and Lattice Dynamics\\
  %Classify using CPC Program Library Subject Index, see (
  % http://cpc.cs.qub.ac.uk/subjectIndex/SUBJECT_index.html)
  %e.g. 4.4 Feynman diagrams, 5 Computer Algebra.
{\em External libraries:} Numpy [1], Scipy [2], Matplotlib [3]\\
  % Fill in if necessary, otherwise leave out.
{\em Subprograms used:}                                       \\
  %Fill in if necessary, otherwise leave out.
{\em Nature of problem:} Materials usually undergo structural phase transitions when the environmental pressure and temperature are elevated to enough high values. The phase transition process obeys the principle of lowest Gibbs free energy. In addition to the static energy, current density functional theory (DFT) calculations can easily give the phonon density of states of lattice vibrations, from which the Helmholtz free energy of phonons are reduced. Then Gibbs free energy can be achieved for the analysis of phase stability and phase transition at high pressure and temperature within the framework of QHA. The problem is to extract the Gibbs free energies from the DFT calculations and automatically analyze the high pressure and temperature phase boundaries among a number of structures.\\
  %Describe the nature of the problem here.
   \\
{\em Solution method:} With the help of numerical interpolation techniques, the Gibbs free energy as a function of pressure/temperature at fixed temperature/pressure can be obtained. Then the QHA based phase boundaries can be automatically determined and plotted by scanning the pressure/temperature at fixed temperature/pressure according to the principle of lowest Gibbs free energy.\\
  %Describe the method solution here.
   \\
{\em Restrictions:} The restriction is from the QHA which takes partially into account the anharmonic effects.\\
  %Describe any restrictions on the complexity of the problem here.
   \\
{\em Unusual features:} The phase boundaries among a number of structures can be automatically determined and plotted, which largely improves the efficiency of phase transition analysis. In addition to some basic thermodynamic properties of each single structure, the Hugoniot pressure-volume-temperature relations are also automatically reduced.\\
  %Describe any unusual features of the program/problem here.
   \\
{\em Additional comments:} This package can treat the phonon density of states data from many packages, such as PHON [4], PHONOPY [5], Quantum ESPRESSO [6], and ABINIT [7].\\
  %Provide any additional comments here.
   \\
{\em Running time:} The examples provided in the distribution take less than a minute to run.\\
  %Give an indication of the typical running time here.

%* Items marked with an asterisk are only required for new versions
\end{small}

\section{Introduction}
The accurate determinations of phase diagrams, equations of state (EOS), and thermodynamic properties of materials are of primary importance to the materials science, the high pressure science, and the geophysics. The high-pressure experimental techniques, including the diamond-anvil cell (DAC) and shock wave (SW) experiments, are main ways to yield such properties precisely. However, except for high costs there are more or less limitations in these experimental techniques, such as pressure or temperature limit and the very short duration time of the compression process in SW experiment during which it is hard to detect phase transitions in real time. So it very urgently needed that accurate theoretical methods are developed to expand the pressure and temperature regime where experiments can not reach.

Current density of functional theory (DFT) has been developed to a powerful tool widely used to calculate and predict all kinds of new properties of materials, and even design new materials theoretically before experimentation~\cite{Curtarolo2013,Zhang2013}. Especially, the DFT has been applied frequently to calculate the high pressure and high temperature (HPHT) dependent properties including the equation of states, thermodynamic properties and phase diagrams under high pressure~\cite{Wu2008,Otero-de-la-Roza2011,Alfe2002,Belonoshko2008}. As for the HPHT dependent properties, the hydrostatic pressure changes subject to the volume changes of the crystal cell provided that the atoms positions are fully relaxed therein. And the inclusion of temperature effects can be realized either by the quasi-harmonic approximations (QHA) or molecular dynamics simulations. The \textit{ab intio} molecular dynamics simulations are computationally expensive. While the QHA is relatively cheap and a good solution to count the temperature effects in materials simulations. Nevertheless, one has to gather the necessary information to extract Gibbs free energies for all volumes considered before the calculations of HPHT properties . And then the related properties and even the phase diagrams can be calculated by numerical techniques. This whole process is complex and time-consuming to finish manually. Especially in the construction of phase diagrams for many structures (more than two), the comparison of their Gibbs free energies and the judgment of stable field for each structure are tough work. This motivates us to design a package named {\sc Phasego} to accomplish all the complicated steps and plot the phase diagrams automatically.

{\sc Phasego} is designed for the easy implementation of phase transition analysis and plot of phase diagrams. It can also calculate the thermodynamic properties of materials, including the thermal expansion coefficients, the bulk moduli, the heat capacities, the thermal pressures, the Gr{\"u}neisen parameters, and the Debye temperatures. For the dynamic response properties of materials, {\sc Phasego} can automatically find the Hugoniot pressure-volume and pressure-temperature relations according to the Rankine-Hugoniot conditions~\cite{Rankine1870,Hugoniot1887}. All these qualities can be obtained based on the QHA by simply preparing a controlling file. More interestingly, based on the QHA the possible phase boundaries of all the structures provided are analyzed and plotted automatically. So the high pressure and temperature phase diagram can be constructed and plotted easily by {\sc Phasego}.

\section{Theoretical backgroud}
\subsection{Quasi harmonic approximation}
The QHA is a phonon-based model of crystal lattice vibrations used to describe volume-dependent thermal effects, such as the thermal expansion. It starts from the assumption that the harmonic approximation holds for every value of the crystal volume, and then takes into account part of anharmonic effects by varying crystal volume.

In the framework of QHA, the Helmholtz free energy of a crystal is written as
\begin{equation}
  F(V,T)=E_{\mathrm{static}} (V)+F_{\mathrm{zp}}(V,T)+F_{\mathrm{ph}}(V,T),
  \label{HelmFreeen}
\end{equation}
where $E_{\mathrm{static}} (V)$ is the first-principles zero-temperature energy of a
static lattice at volume $V$, and $F_{\mathrm{zp}}$ is the zero-point motion energy of the lattice given by
\begin{equation}
  F_{\mathrm{zp}}=\frac{1}{2}\sum_{\mathbf{q},j}\hbar \omega_j(\mathbf{q},V).
  \label{Ezp}
\end{equation}
The last term is the phonon free energy due to lattice vibrations, and can be obtained from
\begin{equation}
  F_{\mathrm{ph}}(V,T)=k_{\mathrm{B}}T \sum_{\mathbf{q},j} \mathrm{ln} \left \{1-\mathrm{exp}\left[-\hbar \omega_j        (\mathbf{q},V)/k_{\mathrm{B}} T\right]\right\},
  \label{Fph}
\end{equation}
where $\omega_j (\mathbf{q},V)$ is the phonon frequency of the $j$th mode of wavevector $\mathbf{q}$ in the first Brillouin zone.

For a metallic material, the Helmholtz free energy at high temperature includes the electronic thermal excitation free energy,
\begin{equation}
  F(V,T)=E_{\mathrm{static}} (V)+ F_{\mathrm{el}}(V,T)+F_{\mathrm{zp}}(V,T)+F_{\mathrm{ph}}(V,T),
  \label{HelmFreeen}
\end{equation}
$F_{\mathrm{el}}$ can be evaluated by the finite temperature DFT with the help of the Fermi-Dirac smearing~\cite{Mermin1965}. At low temperature, $F_{\mathrm{el}}$ is very small and can be neglected, but at high temperature it is large and probably changes final conclusions.

If we write $F_{\mathrm{latt}}$ as
\begin{equation}
F_{\mathrm{latt}} = F_{\mathrm{zp}} + F_{\mathrm{ph}},
\end{equation}
it can also be calculated from the phonon density of states $g(\omega)$ via~\cite{Lee1995}
\begin{equation}
F_{\mathrm{latt}} = \int \left[\frac{1}{2} \hbar \omega + k_{\mathrm{B}}T\mathrm{ln}(2\mathrm{sinh}\frac{\hbar\omega}{2k_{\mathrm{B}}T})\right]\mathrm{d}\omega g(\omega),
\end{equation}
where $k_B$ is the Boltzmann constants.
The phonon density of states is written as
\begin{equation}
g_j(\omega)=\frac{V}{(2\pi)^3}\int \mathrm{d} \mathbf{q} \delta \left[\omega-\omega_j(\mathbf{q})\right].
\end{equation}
And the total densities of states are normalized to,
\begin{equation}
\int \mathrm{g}(\omega)\mathrm{d}\omega=3nN,
\end{equation}
where $n$ is the number of atoms in the unit cell and $N$ is the number of the unit cells.

\subsection{Fitting equation of state}
For a crystal structure, when obtained the Helmholtz free energies at different volumes and fixed temperature, one can derive the analytical function of $F(V)$ at certain temperature by fitting EOS. The types of the EOS in {\sc Phasego} package are: Murnaghan~\cite{Murnaghan1937}, Birch~\cite{Westbrook1995}, 3rd-order Birch-Murnaghan~\cite{Birch1947}, 4rd-order Birch-Murnaghan~\cite{Birch1947}, Vinet~\cite{Hebbache2004}, Vinet Universal~\cite{Vinet1989}, 3rd-order Natural strain~\cite{Poirier1998}, 4rd-order Natural strain~\cite{Poirier1998}, and the 3rd and 4rd Polynomial. Then the pressure $P$ is obtained by
\begin{equation}
P=-\left(\frac{\partial F}{\partial V}\right)_T.
\end{equation}
One can calculate Gibbs free energy $G$ as a function of temperature ($T$) and pressure ($P$) via,
\begin{equation}
G(T,P) = F(V,T) + PV.
\end{equation}

Once the Gibbs free energy is known, the phase stabilities and transitions can be determined by comparing the Gibbs free energies of different structures at fixed pressure or temperature.

\subsection{The derivation of thermodynamic properties}
Other thermodynamic properties can be reduced numerically according to the thermodynamics relations~\cite{Poirier2000}. The thermal pressure $P_{\mathrm{th}}$ at fixed volume is
\begin{equation}
P_{\mathrm{th}} = P(V,T)-P(V,0).
\end{equation}
The enthalpy $H$ does not include thermal effects and can be written as
\begin{equation}
H=E_{\mathrm{static}}(V)+PV.
\end{equation}
The volume thermal expansion coefficient $\alpha$ can be derived via
\begin{equation}
\alpha= \frac{1}{V}\left(\frac{\partial V}{\partial T}\right)_P,
\end{equation}
and the constant temperature bulk modulus is calculated from
\begin{equation}
B_T=-V\left(\frac{\partial P}{\partial V}\right)_T.
\end{equation}
The entropy at constant volume  is
\begin{equation}
S=-\left(\frac{\partial F}{\partial T}\right)_V.
\end{equation}
Then the constant volume heat capacity can be calculated from
\begin{equation}
C_V=T\left(\frac{\partial S}{\partial T}\right)_V =-T\left(\frac{\partial^2 F}{\partial T^2}\right)_V.
\end{equation}
Hence, the thermodynamic Gr{\"u}neisen parameter $\gamma_{\mathrm{th}}$ is derived via
\begin{equation}
\gamma_{\mathrm{th}}=\frac{\alpha B_T V}{C_V}.
\end{equation}
The adiabatic bulk modulus is written as
\begin{equation}
B_S=B_T\left(1+\gamma_{\mathrm{th}}\alpha T\right),
\end{equation}
and the constant pressure heat capacity is calculated from
\begin{equation}
C_P=C_V\left(1+\gamma_{\mathrm{th}}\alpha T\right)=C_V+\alpha^2 B_T VT.
\end{equation}

The Hugoniot pressure-volume and pressure-temperature relations are reduced according to the Rankine-Hugoniot conditions~\cite{Rankine1870,Hugoniot1887},
\begin{equation}
\frac{1}{2}P_H\left(V_0-V_H\right)=\left(E_H-E_0\right),
\label{rh}
\end{equation}
where $E_H$ is the molar internal energy along the Hugoniot, and $E_0$ and $V_0$ are the molar internal energy and volume at zero pressure and room temperature, respectively. Because $P_H$ and $E_H$ are both the functions of temperature, temperature $T$ can be found by numerically solving Eq.\ref{rh} at fixed volume $V_H$.

The internal energy, entropy, and constant volume heat capacity of lattice vibrations can be directly achieved from the phonon density of states (normalized to $3nN$) via~\cite{Lee1995},

\begin{equation}
    E=\frac{\hbar}{2}\int^{\omega_{max}}_0\omega\mathrm{coth}\frac{\hbar\omega}{2k_\mathrm{B}T}g(\omega)\mathrm{d}\omega,
\end{equation}

\begin{equation}
    S=k_\mathrm{B}\int^{\omega_{max}}_0\left[\frac{\hbar\omega}{2k_\mathrm{B}T}\mathrm{coth}\frac{\hbar\omega}{2k_\mathrm{B}T}-\mathrm{ln}\left(2\mathrm{sinh}\frac{\hbar\omega}{2k_\mathrm{B}T}\right)\right]g(\omega)\mathrm{d}\omega,
\end{equation}
and
\begin{equation}
    C_V=k_\mathrm{B}\int^{\omega_{max}}_0\left(\frac{\hbar\omega}{2k_\mathrm{B}T}\right)^2\mathrm{csch}^2\left(\frac{\hbar\omega}{2k_\mathrm{B}T}\right)g(\omega)\mathrm{d}\omega,
    \label{cv}
\end{equation}
respectively.

%    S=3nNk_\mathrm{B}\int\left[\frac{\hbar\omega}{2k_\mathrm{B}T}\mathrm{coth}

According to the Debye approximation, the specific heat can be expressed as,
\begin{equation}
    C_V = 9N_Ak_\mathrm{B}\left(\frac{T}{\Theta_D}\right)^3 \int ^{\Theta_D/T}_0 \frac{x^4e^x}{(e^x-1)^2}\mathrm{d}x
    \label{thd}
\end{equation}
where $\Theta_D$ is Debye temperature and $N_A$ is the Avogadro constant. So, at a fixed volume the Debye temperature $\Theta_D$ at each temperature $T$ can be obtained from Eqs.~\ref{cv} and~\ref{thd}.

\section{Description of the package and Input/output files}

\subsection{Installation requirements}
The installation of {\sc Phasego} package is very easy. {\sc Phasego} is based on Python, but to increase the computation speed and realize the numerical interpolation and extrapolation, it also uses Numpy and Scipy python libraries.

The following packages are required:\\
$\bullet$ Python 2.6 or later.\\
$\bullet$ NumPy.\\
$\bullet$ Scipy.\\
$\bullet$ Matplotlib.\\

Matplotlib is used for the automatic plot of the HPHT phase diagram. In the Ubuntu system, just simply execute {\tt sudo apt-get install python python-numpy python-scipy python-matplotlib}. In the Windows systems, one can install {\tt pythonxy} package for all the necessary libraries.

\subsection{Installation}
There are two methods to install {\sc Phasego}. When one has the python setuptools module installed, the first method is to execute {\tt python setup.py install} in the {\sc Phasego} root directory as the administrator. As a common user, one can also specify a directory to install using additional ``{\tt --prefix=/\\path/to/install/}". If one has no python setuptools, then the second method to install is: first put the following {\tt PATH} and {\tt PYTHONPATH} environment variable in user's {\tt $\sim$/.bashrc} file:\\

{\tt
export PATH=\$PATH:/path/to/phasego/src
 
export PYTHONPATH=\$PYTHONPATH:/path/to/phasego/src}
\\

or in {\tt $\sim$/.cshrc} file:\\

{\tt
setenv PYTHONPATH /path/to/phasego/src:\${PYTHONPATH}

setenv PATH /path/to/phasego/src:\${PATH}}
\\

Then, execute {\tt chmod +x /path/to/phasego/src/phasego}. Finally execute {\tt source $\sim$/.bashrc} or {\tt source $\sim$/.cshrc}.

\subsection{Run}
Just type {\tt phasego} to run the phase analysis in your work directory contained a controlling file {\tt phasego.in}. All the controlling arguments are detailed below.

\subsection{Input files}
The input files needed by {\sc Phasego} package are the static volume-energy data files and the phonon density of states files for each volume. The structure information can be obtained from the crystal structure prediction codes, such as our recently developed {\sc Muse} package~\cite{Liu2014} and the {\sc calypso} package~\cite{wang2010}. All the files are placed in the {\tt inp-phasename} directory for each single structure, where {\tt phasename} is the name of this structure named by the user in the controlling file {\tt phasego.in}. The common prefix name of the static volume-energy data files for all structures are specified in the {\tt phasego.in} file, e.g., {\tt ve-}. The full file names of volume-energy data files are this prefix name plus ``{\tt T}", where {\tt T} is the temperature at which the volume-energy data are calculated. To include the electronic thermal excitation effects (especially for metallic materials), the energies can be calculated within the framework of finite temperature DFT. There should be a number of volume-energy data files, e.g., {\tt ve-0}, {\tt ve-50}, {\tt ve-100}, {\tt ve-150}, and so on. For nonmetallic materials, {\tt T} equals to 0 and there is only one file, e.g., {\tt ve-0}. The phonon dos file is named as {\tt ph.dos-volume}, where {\tt volume} is exactly the same value in the volume-energy file. {\tt ph.dos-} is the common prefix name specified in {\tt phasego.in} for all the structures. The density of states is normalized to $3nN$, where $n$ is the number of atoms in the unit cell and $N$ is the number of the unit cells. The frequencies are in cm$^{-1}$ and dos are in states/cm$^{-1}$.

The following input file example is for the phase transition analysis of GaN at HPHT. The main input file is named as {\tt phasego.in}, in which the lines started with `{\tt \#}' are neglected. Each argument and its value(s) are placed in the same line, in any order and with any number of blank lines. The values of each argument are separated by whitespace and placed after `{\tt =}'. The arguments are all self-explained in their comment line(s).\\

\begin{flushleft}
\tt {
\#\#\#=========================================================\#\#\#\\
\#\#\#=================== Phasego input file ==================\#\#\#\\
\#\#\#=========================================================\#\#\#\\

\#\#\# Lines started with "\#" are neglected.\\
\#\#\# Parameter and value(s) are placed in the same line, in any \\
\#\#\# order with any number of blank lines.\\
\#\#\#\\
\# Note:\\
\#\#\#\\
\# A. volumes in Bohr$^3$, energies in Ry.\\
\# B. Num\_atoms is the number of atoms in the unit cell for \\
\# $  $ $ $ $ $ energy and phonon dos calculations.\\
\# C. dos files: normalized to 3*N, and frequencies in cm-1 \\
\# $  $ $ $ $ $ and dos in states/cm-1.\\
\#\\
\# EOS Names (number):\\
\# 1. Murnaghan: Murnaghan EOS (F. D. Murnaghan, Am. J. Math.\\
\# $  $ $ $ $ $ $ $ 49, 235 (1937))\\
\# 2. Birch: Birch EOS (From Intermetallic compounds:\\
\# $  $ $ $ $ $ $ $ Principles and Practice , Vol I: Principles. pages \\
\# $  $ $ $ $ $ $ $ 195-210)\\
\# 3. BirchMurnaghan: Birch-Murnaghan 3rd-order EOS (F. Birch,\\
\# $  $ $ $ $ $ $ $ Phys. Rev. 71, p809 (1947))\\
\# 4. BirchMurnaghan: Birch-Murnaghan 4rd-order EOS (F. Birch,\\
\# $  $ $ $ $ $ $ $ Phys. Rev. 71, p809 (1947))\\
\# 5. Vinet: Vinet EOS (Vinet equation from PRB 70, 224107)\\
\# 6. Universal: Universal EOS (P. Vinet et al., J. Phys.:\\
\# $  $ $ $ $ $ $ $ Condens. Matter 1, p1941 (1989))\\
\# 7. Natural strain 3rd-order EOS (Poirier J-P and Tarantola\\
 \# $  $ $ $ $ $ $ $ A,Phys. Earth Planet Int. 109, p1 (1998))\\
\# 8. Natural strain 4rd-order EOS (Poirier J-P and Tarantola \\
\# $  $ $ $ $ $ $ $ A,Phys. Earth Planet Int. 109, p1 (1998))\\
\# 9. Cubic polynomial\\
\#10. 4th polynomial\\
\#\#\#===========================================================\#\#\#\\
$ $\\

\# The name of EOS used for fitting\\
Eos\_Name = 1\\
$ $\\

\# The user-defined name for each corresponding structure. The \\
\# ve and phonon dos files are placed in each inp-name dir. The \\
\# output files are placed in each out-name dir.\\
Names\_of\_Strs = rocksalt wurtzite zencblende \\
$ $\\

\# Number of atoms used for ve and phonon dos data of each str.\\
Num\_atoms = 2 4 2 \\
$ $\\

\# The prefix name of volume-energy data files for each str.\\
\# The full name is the prefix name plus "T", where T is \\ 
\# temperature to take into account the electronic thermal \\ 
\# excitation by finite temperature DFT. If including the \\
\# electronic thermal excitation free energy, the temperature \\
\# start, end, and interval should be the same as Tdata below.\\
VE\_data\_File\_Name = ve-\\
$ $\\

\# Phonon dos file prefix name, i.e., it plus the volume value \\
\# in ve file is the full name.\\
Ph\_Dos\_File\_Base\_Name = ph.dos-\\
$ $\\

\# Temperature data (K): start, end, interval\\
Tdata = 0 6000 100\\
$ $\\

\# Pressure data (GPa): start, end, interval\\
Pdata = 0 40 1\\
$ $\\

\# If calculate thermal properties, yes or no\\
If\_Incl\_Phonon = yes\\
$ $\\

\# If include electronic thermal excitation free energy by \\
\# finite temperature DFT, for metals\\
If\_Incl\_Electronic\_Excitation = no\\
$ $ \\

\# If analyze the potential phase transition P-T points between \\
\# these strs, yes or no\\
If\_Analysize\_Phase\_Transition = yes\\
$ $\\

\# If plot phase diagram using mathplotlib, yes or no\\
If\_Plot = yes\\
$ $\\

\# If analyze Hugoniot PTV, yes or no\\
If\_Hugoniot = yes\\
$ $\\

\# If calculate Debye Temperature, yes or no
$ $\\
If\_Calc\_Debye\_Temp = yes\\

}

\end{flushleft}

\subsection{Output files}
The output files for single phase analysis are placed in the {\tt out-phasename} directory of each structure. The {\tt Alpha.dat} file contains the thermal expansion coefficients data as the function of temperature at fixed pressures. The {\tt B\_S.dat} file has the adiabatic bulk modulus data as the function of temperature at fixed pressures. The {\tt B\_T\_P.dat} and {\tt B\_T\_T.dat} files contain the isothermal bulk moduli at fixed pressures and temperatures, respectively. The {\tt C\_P.dat} file contains the outcomes of heat capacities at fixed pressures. The {\tt C\_V\_P.dat} and {\tt C\_V\_V.dat} files have the constant volume heat capacities at fixed pressures and volumes, respectively. The {\tt DebyeT.dat} file has the Debye temperatures as the function of temperature at fixed volumes. The {\tt Enthalpy.dat} is the data file of the enthalpy as a function of pressure. The {\tt Entropy\_V.dat} and {\tt Entropy\_P.dat} collect the total entropies at fixed volumes and pressures, respectively. The {\tt fittedC\_V\_V.dat} includes the fitted $C_V$ at fixed volumes. The {\tt fittedE\_V.dat} is the data file of the fitted internal energy and pressure. The {\tt FittedHelmFreeE\_T.dat} file contains the fitted Helmholtz free energies at fixed temperatures. The {\tt gamma\_P.dat} has the thermodynamic Gr\"uneisen parameters as the function of temperature at fixed pressures. The {\tt G\_P.dat} and {\tt G\_T.dat} files include the Gibbs free energies as functions of temperature and pressure, respectively. The {\tt HugoniotPTV.dat} is the output file of the Hugoniot P-T-V data. The {\tt PV\_T.dat} file contains the pressure-volume data at fixed temperatures. The {\tt ThermalP\_T.dat} and {\tt ThermalP\_V.dat} have the thermal pressure data at fixed pressures and volumes, respectively. The {\tt VT\_P.dat} is the output file of the volume-temperature data at fixed pressures. The {\tt C\_V-ph-direct.dat}, {\tt Entropy-ph\_V.dat} and {\tt E-ph\_V.dat} are the constant volume heat capacities, entropy and internal energy of lattice vibrations directly from the phonon density of states, respectively. 

The phase transition data and the automatically plotted phase diagram figures are placed in the {\tt Phase-PT} directory. The {\tt P-T.dat} file contains the transition pressures at fixed temperatures, which are obtained by canning pressure at each fixed temperature. Similarly, the {\tt T-P.dat} file contains the transition temperatures at fixed pressures obtained by scanning temperature at each fixed pressure. They can be plotted by other plot softwares. The phase boundaries are indexed by ``$|$" and the phase transitions are labelled by "------$>$".

\section{Two examples: phase analyses of GaN and Fe}
To test the ability of {\sc Phasego} package, I performed phase analyses for the semiconductor GaN and the metal Fe. {\sc Phasego} can automatically determine the phase boundaries and plot the phase diagrams for the several competing structures. The total time to run is less than a minute both for the two cases. For GaN, the thermodynamic properties of the wurtzite phase are also presented and discussed below.
 
%%%%%%%%% Begin %%%%%%%%%%
\begin{figure}[h]
\centering
\includegraphics[width=6.5cm]{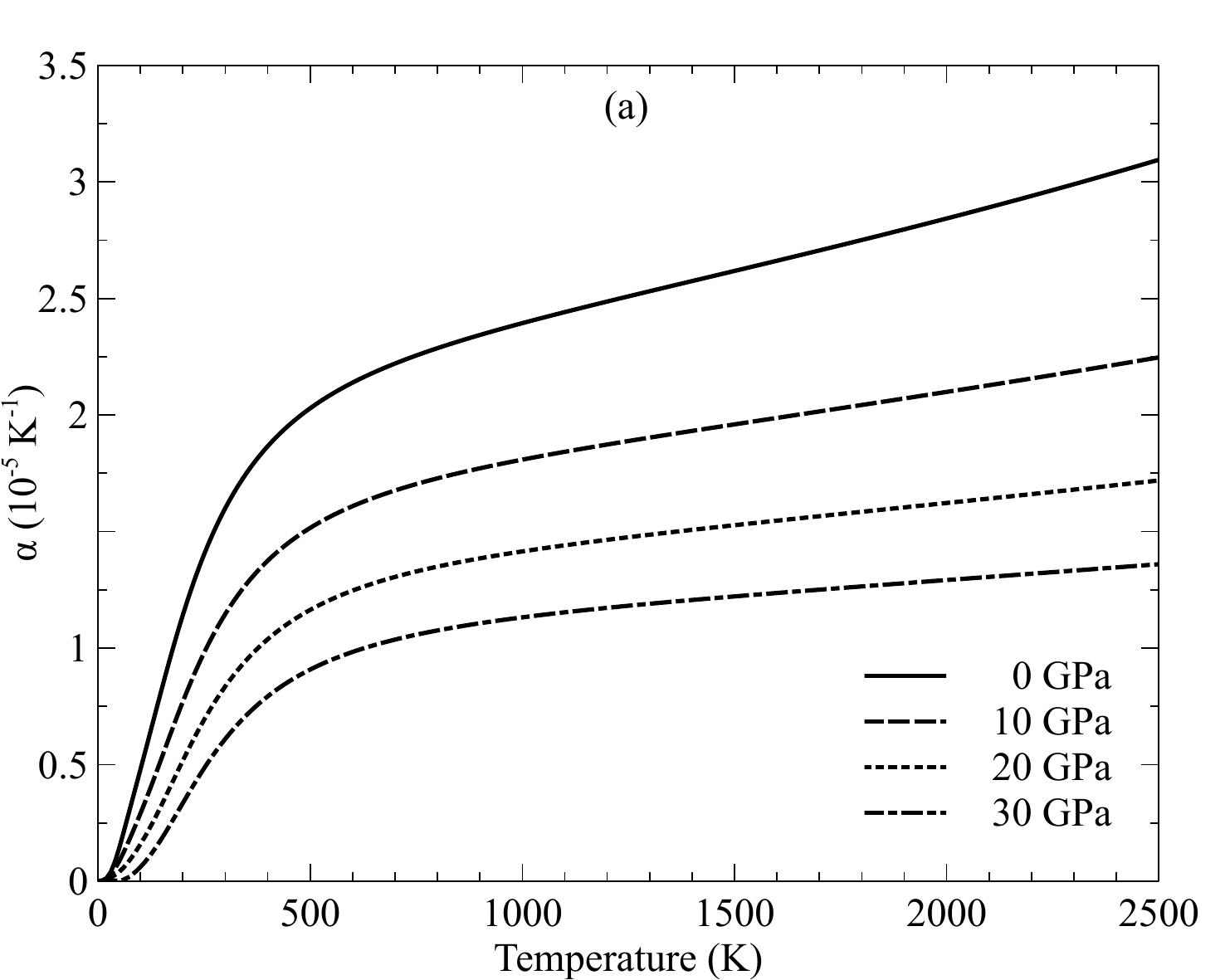}
\includegraphics[width=6.5cm]{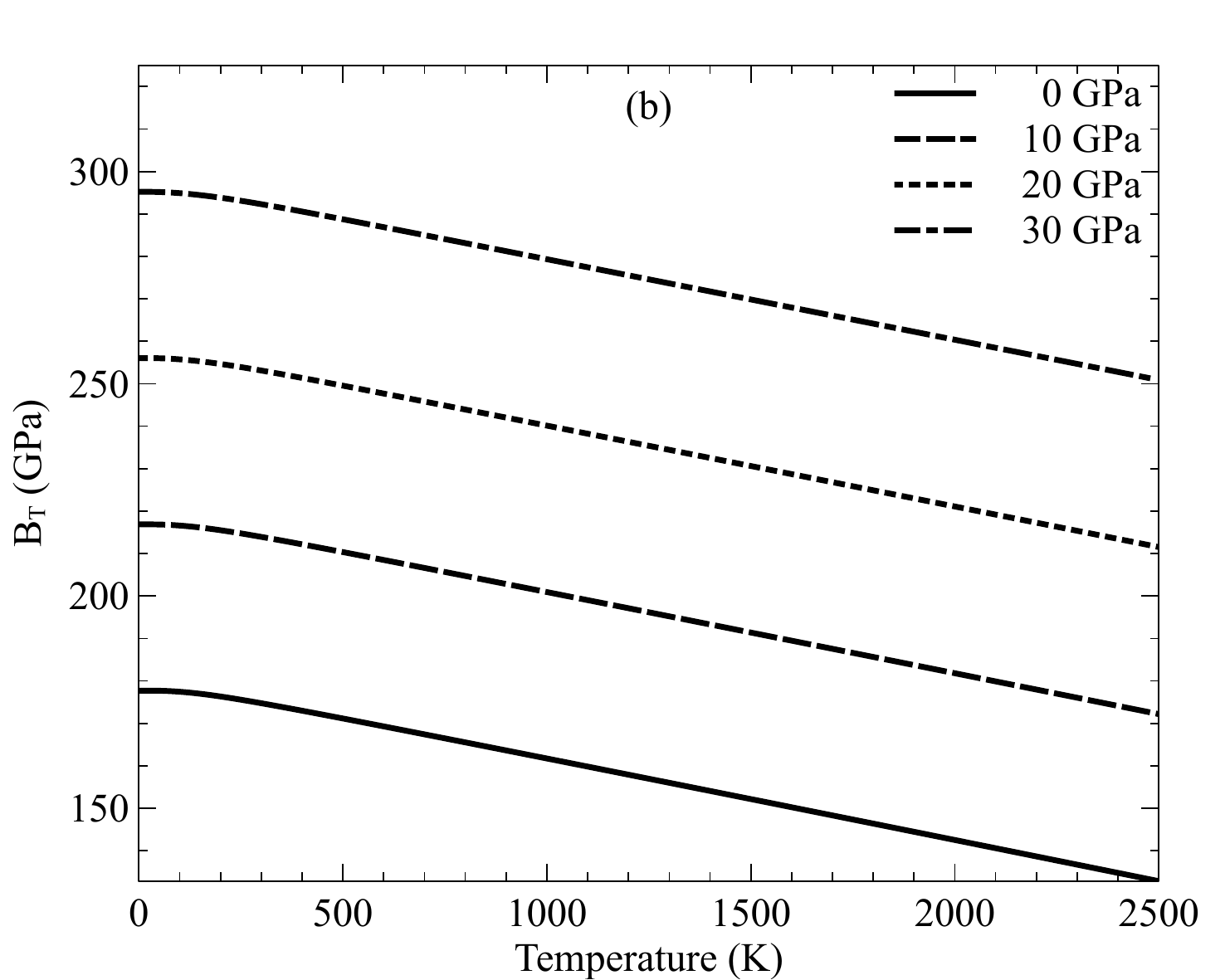}
\caption{(a) The thermal expansion coefficients of wurtzite-GaN as the function of temperature at different high pressures. (b) The isothermal bulk moduli of wurtzite-GaN as the function of temperature at different pressures.}
\label{fig1}
\end{figure}
%%%%%%%%%% End %%%%%%%%%%%

Their phonon frequencies and phonon density of state were calculated using the density functional perturbation theory (DFPT)~\cite{Baroni1987,Baroni2001}, as implemented in the QUANTUM-ESPRESSO package~\cite{pwscfcite}. The exchange-correlation functional used is generalized gradient approximation (GGA) parametrized by PBE~\cite{Perdew1996} and the pseudopotential is ultrasoft pseudopotential~\cite{Vanderbilt1990}. The Fe pseudopotential is the newest version from the recently developed pseudopotential library~\cite{Garrity2014}. We careful tested on {\bf k} and {\bf q} grids, the kinetic energy cutoff, and other technical parameters to ensure good convergence of phonon frequencies. The kinetic energy cutoff and the energy cutoff for the electron density were chosen to be 50 Ryd., 500 Ryd for GaN, and 40 Ryd. and 400 Ryd. for Fe in both total energy and  phonon dispersion calculations, respectively. We applied the Marzari-Vanderbilt~\cite{Marzari1999} smearing method with the smearing width of 0.03 Ryd. For the electronic thermal excitation in Fe, the smearing with the Fermi-Dirac function was applied to set the electronic temperature.
 
%%%%%%%%% Begin %%%%%%%%%%
\begin{figure}[h]
\centering
\includegraphics[width=6.5cm]{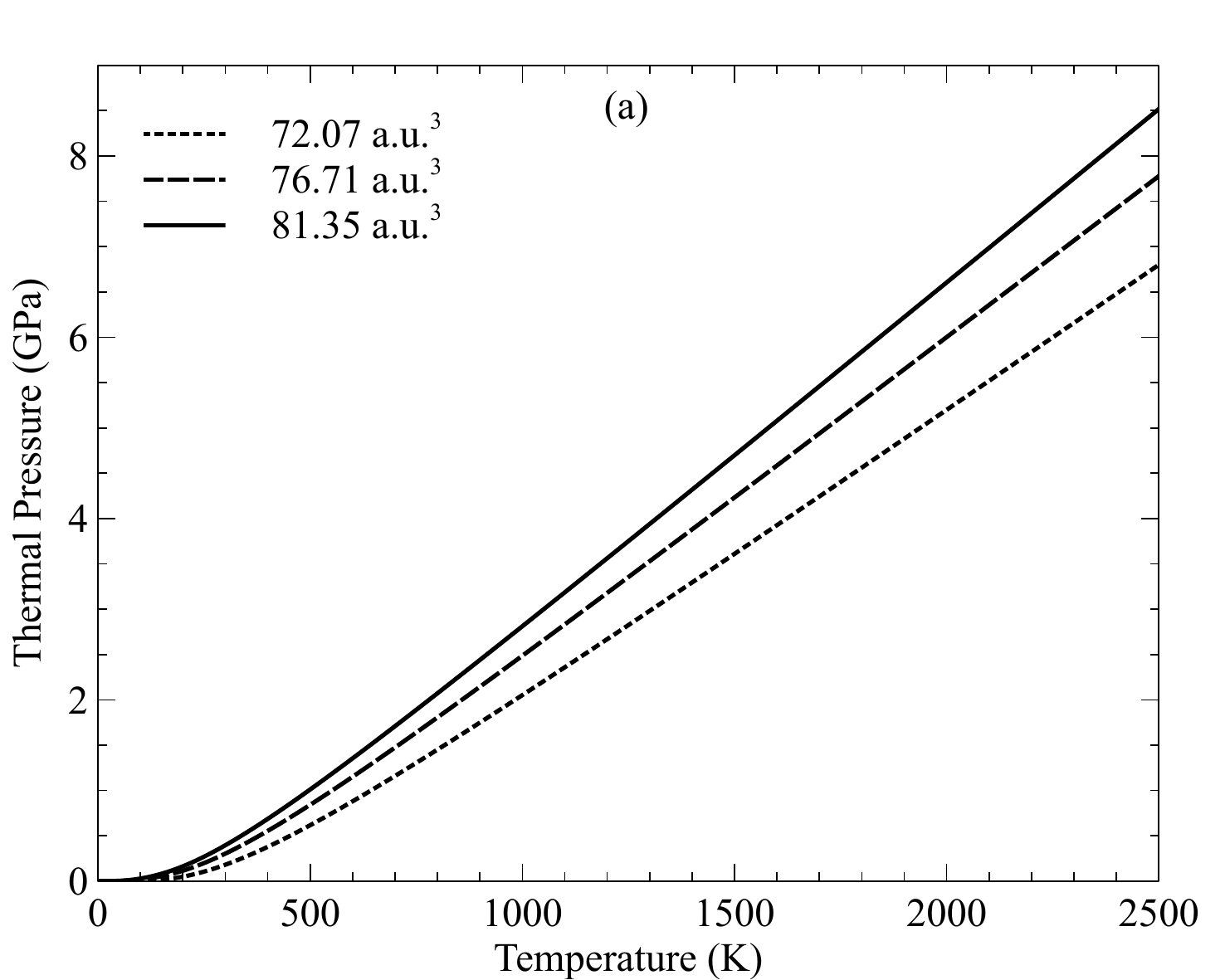}
\includegraphics[width=6.5cm]{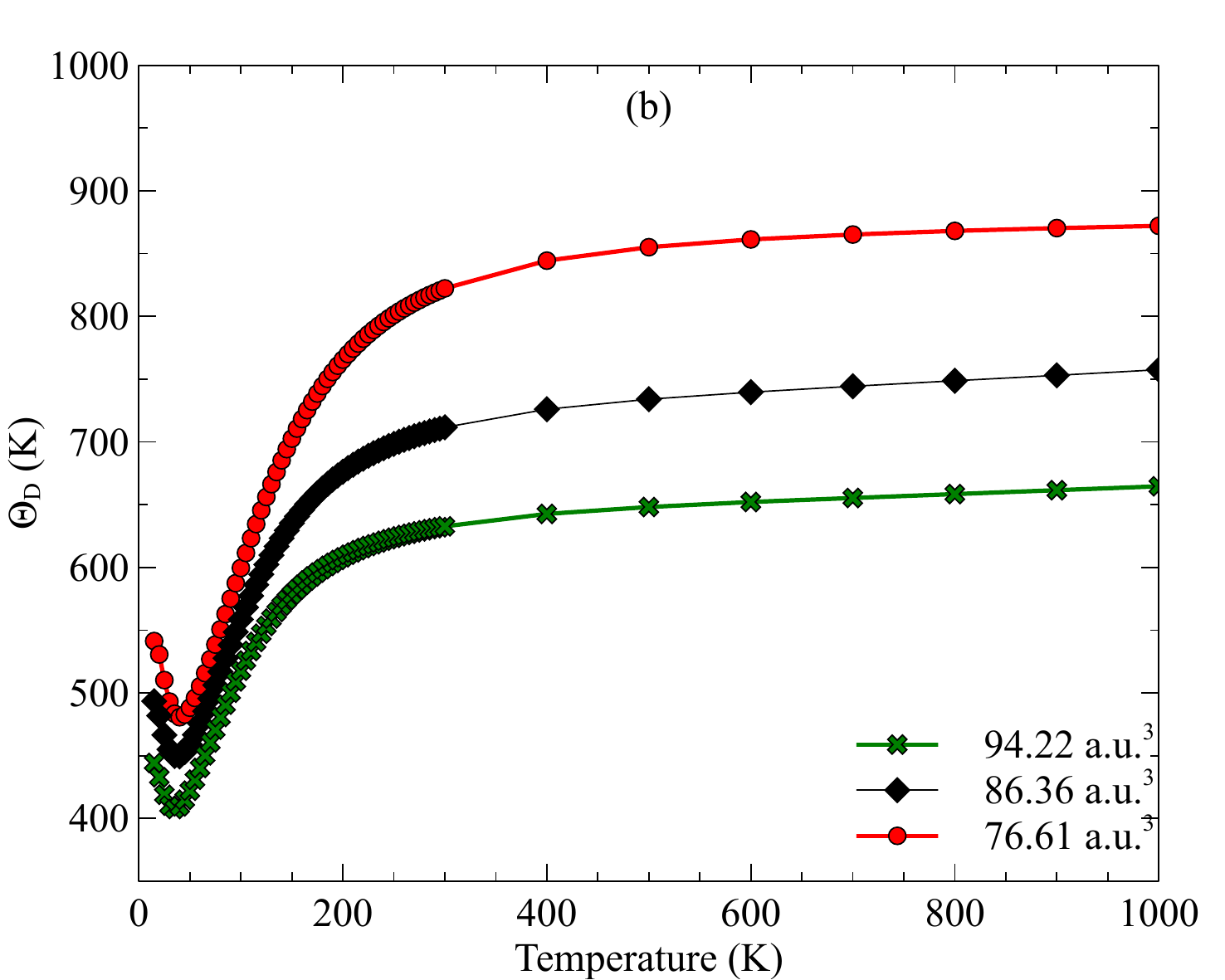}
\caption{(a) The thermal pressure of wurtzite-GaN as the function of temperature at different pressures. (b) The Debye temperature of wurtzite-GaN as the function of temperature at fixed volumes.}
\label{fig2}
\end{figure}
%%%%%%%%%% End %%%%%%%%%%%

\subsection{Thermodynamics of GaN}
The thermodynamic properties of its zinc blende, wurtzite and rocksalt structures are first calculated before phase diagrams calculations. Figure\ref{fig1}a shows the thermal expansion coefficients of wurtzite-GaN as the function of temperature at fixed pressures 0, 10, 20, and 30 GPa. The thermal expansion coefficient increases with temperature at fixed pressure. But it decreases with pressure at fixed temperature. The isothermal bulk modulus as a function of temperature at fixed pressure is illustrated in Fig.\ref{fig1}b, which shows its decreasing behaviour with temperature at fixed pressure.
%%%%%%%%% Begin %%%%%%%%%%
\begin{figure}[h]
\centering
\includegraphics[width=6.5cm]{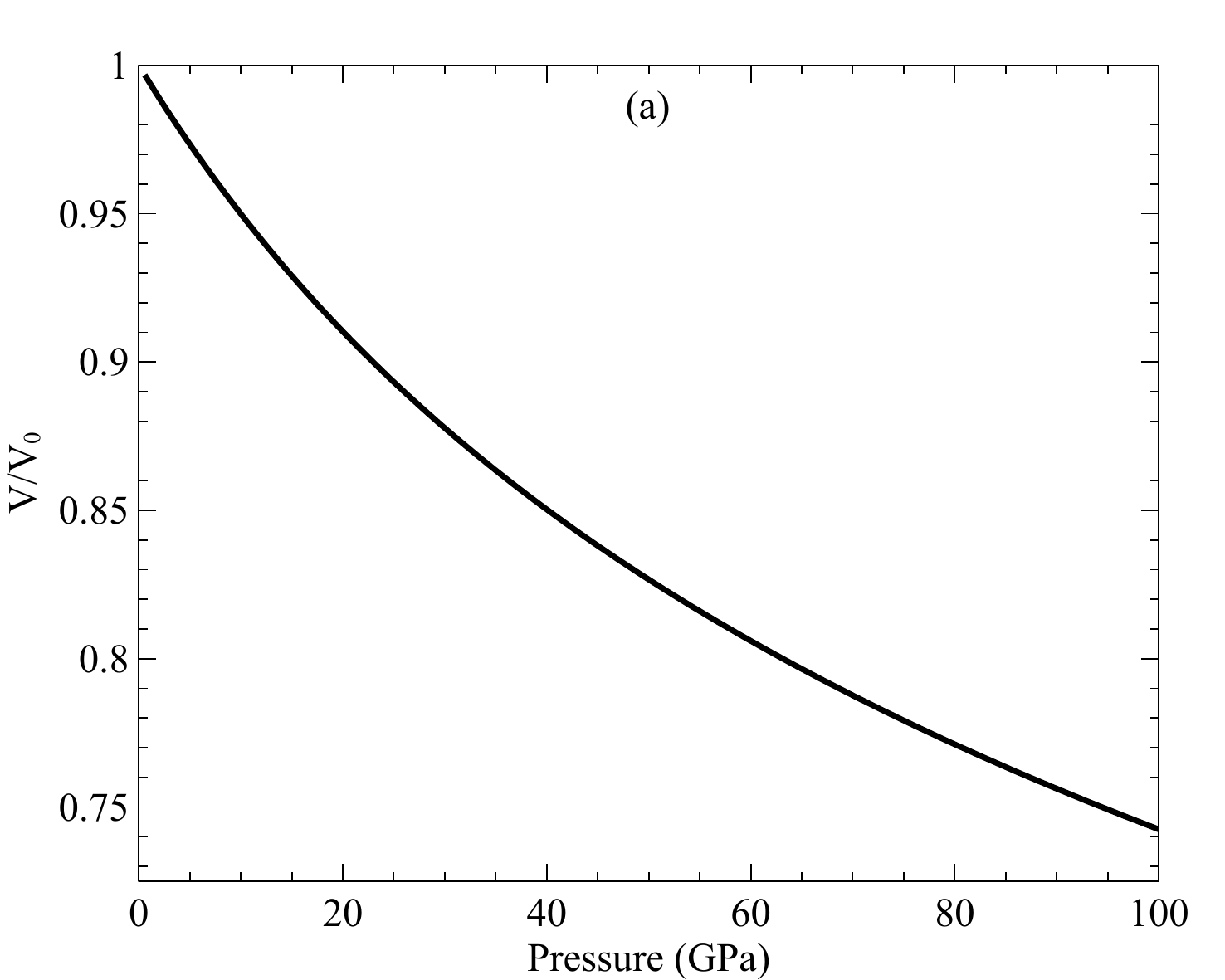}
\includegraphics[width=6.5cm]{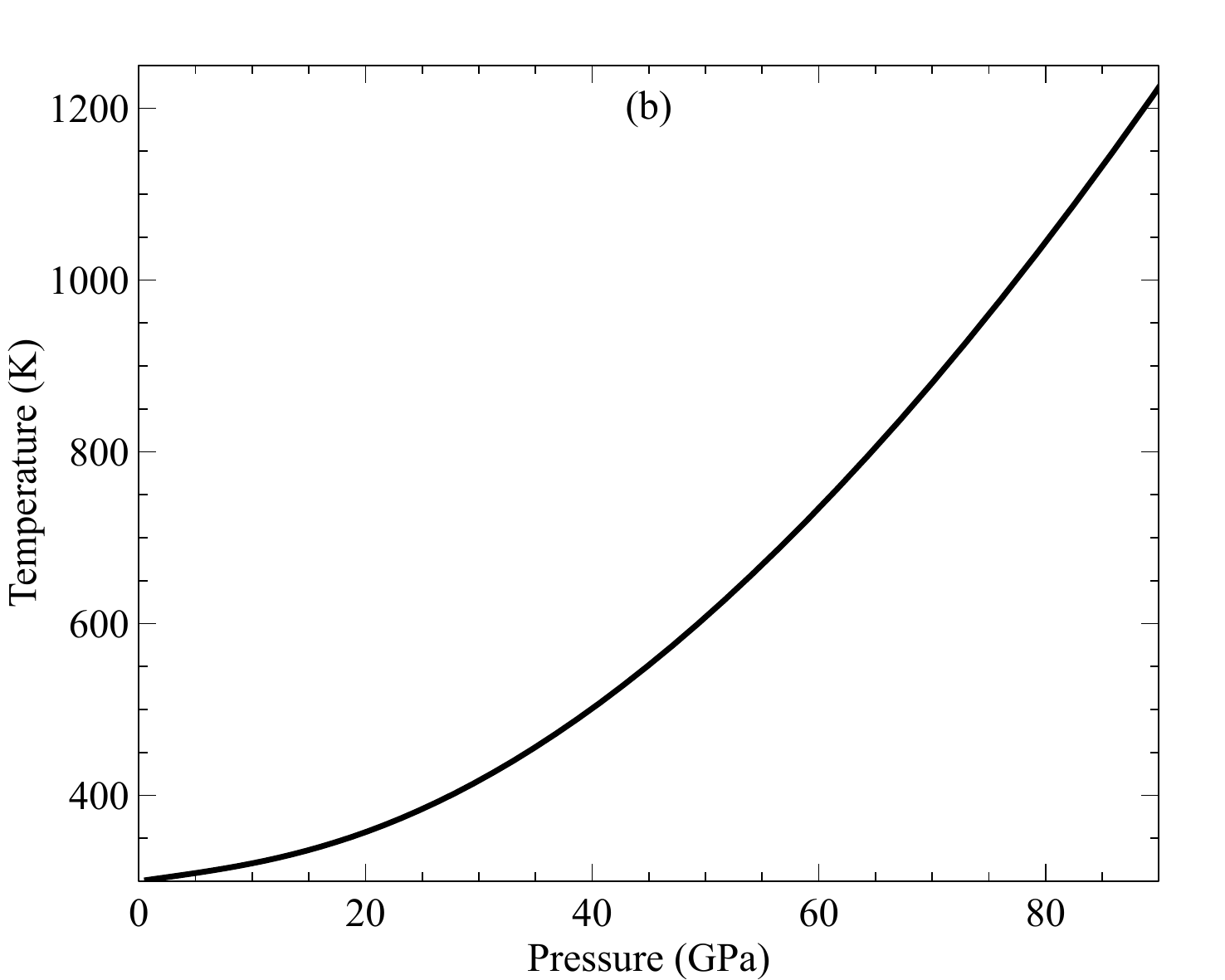}
\caption{(a) The Hugoniot volume versus pressure relation of wurtzite-GaN from single phase analysis. (b) The Hugoniot temperature versus pressure relation of wurtzite-GaN.}
\label{fig3}
\end{figure}
%%%%%%%%%% End %%%%%%%%%%%

The thermal pressure of wurtzite-GaN as a function of temperature at fixed atomic volumes 72.07, 76.71, and 81.35 a.u.$^3$ are reduced and shown in Fig.\ref{fig2}a. The thermal pressure increases with temperature as fixed volume. But unusually the thermal pressure of wurtzite-GaN decreases with decreasing volume. The Debye temperatures of wurtzite-GaN (Fig.\ref{fig2}b) exhibit dramatic drops before 50 K and then increase quickly before 300 K, and finally converge to constants at fixed volumes.

Manually reducing the Hugoniot pressure-temperature-volume relations is a time-consuming work. For the {\sc Phasego} package, it is very easy to implement. As an example, the Hugoniot pressure-volume and pressure-temperature relations of wurtzite-GaN are calculated and presented in Fig.\ref{fig3}a.

%%%%%%%%% Begin %%%%%%%%%%
\begin{figure}[h]
\centering
\includegraphics[width=6cm]{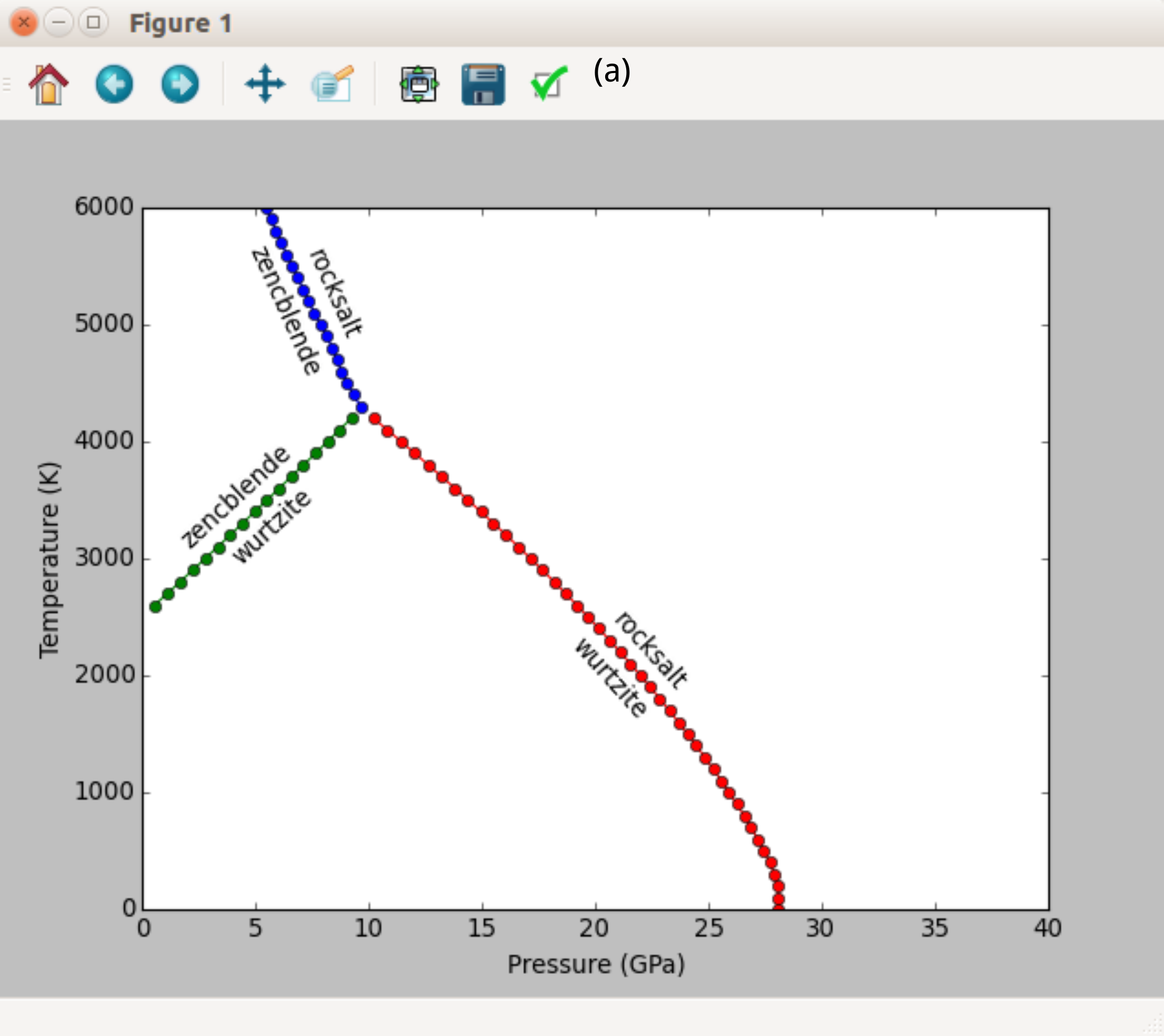}
\includegraphics[width=7cm]{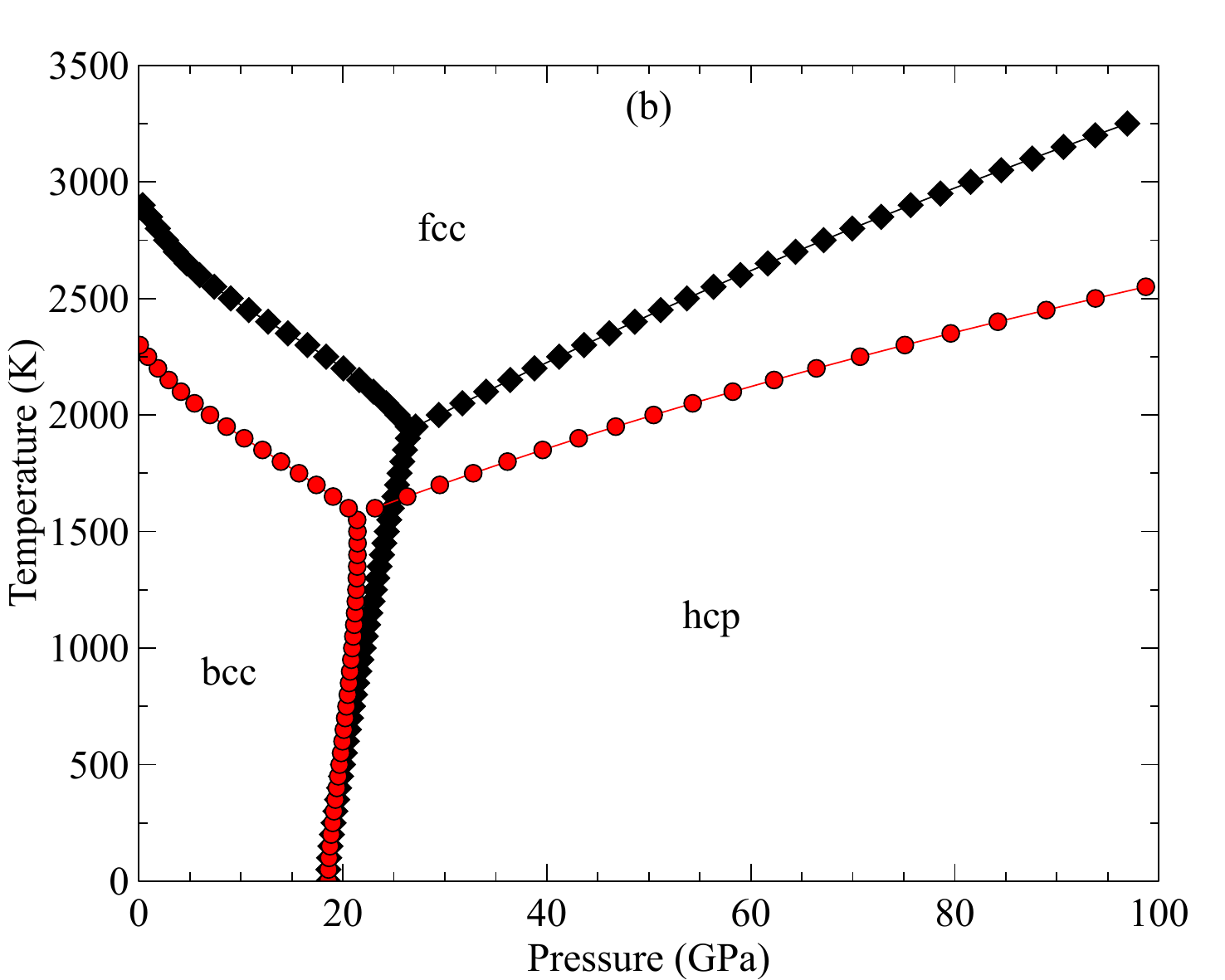}
\caption{(a) The QHA based high pressure and temperature phase diagram of GaN from the automatic analysis and plot of phase transition. (b) The QHA based high PT phase diagram of Fe with (red symbols) and without (black symbols) electronic thermal excitation free energy corrections.}
\label{fig4}
\end{figure}
%%%%%%%%%% End %%%%%%%%%%%

\subsection{Phase diagrams of GaN and Fe}
Figure~\ref{fig4}a is the automatic plot of the phase diagram of GaN. One only needs provide the necessary volume-energy and phonon density of state data for different structures covering the pressure range of interest. The {\sc Phasego} package will gather the Gibbs free energy information and judge the stable field of each structure, and finally plot the phase diagram and label the phase boundary between two structures automatically. The triple point is also clearly shown in the phase diagram. For metallic materials, the electronic thermal excitation free energies at high temperature can not be neglected. In the case of Fe, the results of automatic phase transition analyses with and without electronic thermal excitation free energy corrections are shown in Fig.\ref{fig4}b. One notes that the electronic thermal excitation free energies alter the phase boundaries much obviously at high temperature.

\section{Conclusions}
\label{conclusions}
I described here the {\sc Phasego} package which implements the automatic calculation and plot of phase diagram based on the QHA. The QHA theory and the numerical derivation process of the thermodynamic properties are also detailed. In order to test the validity and efficiency of the {\sc Phasego} package, I performed the phase transition analyses and thermodynamic properties calculation for GaN. For the metallic materials, I took Fe as an example, and found the electronic thermal excitation free energy can not be neglected at high temperature and it changes the phase diagram of Fe obviously.

\section{ACKNOWLEDGMENTS}
The research was supported by the National Natural Science Foundation of China (11104127, 11104227), the NSAF of China under grant No. U1230201/A06, the Project 2010A0101001 funded by CAEP, the Program for Innovative Research Team in Henan University of Science and Technology under Grant No. 13RTSTHN020, and the Science Research Scheme of Henan Education Department under Grand No. 2011A140019.

\label{}

%% The Appendices part is started with the command \appendix;
%% appendix sections are then done as normal sections
%% \appendix

%% \section{}
%% \label{}

%% References
%%
%% Following citation commands can be used in the body text:
%% Usage of \cite is as follows:
%%   \cite{key}         ==>>  [#]
%%   \cite[chap. 2]{key} ==>> [#, chap. 2]
%%

%% References with bibTeX database:

\bibliographystyle{elsarticle-num}
%\bibliography{Refs}

%% Authors are advised to submit their bibtex database files. They are
%% requested to list a bibtex style file in the manuscript if they do
%% not want to use elsarticle-num.bst.

%% References without bibTeX database:

% \begin{thebibliography}{00}

%% \bibitem must have the following form:
%%   \bibitem{key}...
%%

% \bibitem{}

% \end{thebibliography}

\end{document}